\documentclass[12pt]{iopart}

\usepackage{amssymb}
\usepackage{amsfonts, amssymb}
\usepackage{mathrsfs}

\def\ear{\end{eqnarray}}
\def\beq{\begin{equation}}             \def\earn{\nonumber \end{eqnarray}}
\def\eeq{\end{equation}}               
\def\bear{\begin{eqnarray}}

\begin{document}

\title[Black holes and wormholes in $f(R)$ gravity with a kinetic curvature scalar]{Black holes and wormholes in $f(R)$ gravity \\ with a kinetic curvature scalar}

\author{Sergey V. Chervon}
\address{Laboratory of  gravitation, cosmology, astrophysics, Ulyanovsk State Pedagogical University, Lenin's Square 4/5, Ulyanovsk, 432071, Russia}
\address{Bauman Moscow State Technical University, 2-nd Baumanskaya street, 5, Moscow, 105005, Russia}
\address{Institute of Physics, Kazan Federal University, Kremlevskaya street 18, Kazan, 420008, Russia}
\ead{chervon.sergey@gmail.com}
\author{J\'ulio C. Fabris}
\address{Nucleo Cosmo-ufes \& Departamento de F\'{\i}sica, Universidade Federal do Esp\'{\i}rito Santo (UFES)
Av. Fernando Ferrari, 514, CEP 29.075-910, Vit\'oria, ES, Brazil}
\address{National Research Nuclear University MEPhI, Kashirskoe sh. 31, Moscow 115409, Russia}
\ead{julio.fabris@cosmo-ufes.org}
\author{Igor V. Fomin}
\address{Bauman Moscow State Technical University, 2-nd Baumanskaya street, 5, Moscow, 105005, Russia}
\ead{ingvor@inbox.ru}
\vspace{10pt}
\begin{indented}
\item[] 26 July 2020
\end{indented}

\begin{abstract}

We study the chiral self-gravitating model (CSGM) of a special type in the spherically symmetric static spacetime in Einstein frame. Such CSGM is derived, by virtue of Weyl conformal transformation, from a gravity model in the Jordan frame corresponding to a modified $f(R)$ gravity with a kinetic scalar curvature.

 We investigate the model using harmonic coordinates and consider a special case of the scaling transformation from the Jordan frame. We find classes of solutions corresponding to a zero potential and we investigate horizons, centers and the asymptotic behavior of the obtained solutions. Other classes of solutions (for the potential not equal to zero) are found using a special relation (ansatz) between the metric components. Investigations of horizons, centers and asymptotic behavior of obtained solutions for this new case are performed as well. Comparative analysis with similar solutions obtained earlier in literature is made.
\end{abstract}

%

\section{Introduction}

An important area of research in astrophysics and cosmology is currently
the study of the geometry of black holes, objects whose existence has been confirmed by the detection of gravitational waves from binary black hole systems reported by LIGO and VIRGO collaborations
\cite{Abbott:2016blz,Abbott-2019}
 and the detection of the shadow of a supermassive black hole reported by Event
Horizon Telescope Collaboration \cite{Akiyama-2019}. Both the detection of gravitational waves and the confirmation of the existence of a shadow of a black hole
seem to be in agreement with the predictions of GR. However, this does not exclude
an alternatives configurations coming from a modified theory of gravity.

 To date, acceleration in the expansion of the universe has been reliably proven
 by various observations such as measurements from supernovae \cite{cfo-Perlmutter:1998np}, \cite{cfo-Riess:1998cb}, cosmic microwave background (CMB) radiation \cite{Ade:2015xua}, \cite{BICEP2-2015}, \cite{WMAP2013}, large scale structure \cite{Seljak:2004xh}, baryon acoustic oscillation \cite{Eisenstein:2005su} and weak lensing \cite{Jain:2003tba}.
 An important direction to justify the observed acceleration of the Universe is the idea of attracting and developing such (modified) theories of gravity, which lead to a geometric justification of acceleration on a very large scale. For the latest reviews of modified gravity theories, see, for example,\cite{Nojiri:2010wj}, \cite{Nojiri:2017ncd}, \cite{Joyce:2014kja} and the fundamental work \cite{Clifton:2011jh}.

The study of space-time static singular solutions on the basis of Einstein gravity and modified gravity theories is relevant in the context of astrophysical compact objects, including black holes \cite{Visser,Lobo,Bronnikov,Bronnikov2,Lobo2,Alexeev:1999zs,Vlasov:2015rna}.
The black holes and wormholes solutions and different types of a scalar fields are considered in a lot of works on the basis of GR and its modifications (see, for example, in \cite{Korolev,Sushkov,Egorov,Bronnikov3}).

Corrections of higher order curvature to the Einstein-Hilbert action arise
when quantum effects are considered in the low-energy limit of
string theories, superstrings, and supergravity needed to build quantum theory
gravity \cite{cfo-Baumann:2014nda}.
 The example of quantum corrections application was demonstrated
  by A. Starobinsky \cite{cfo-Starobinsky:1980te} in cosmology. It was shown that such corrections may control an accelerated expansion of the Universe at its early stage of evolution (inflation).
  This kind of models has been developed considering 6th-order corrections in theories of gravity
  of the kind  $R+\alpha R^2  +\gamma R\opensquare R$, where $\alpha$ and $\gamma$ are some constants,
  and the additional terms that modify the Einstein theory were, with the aid of conformal transformations
 of the metric, put into correspondence to certain effective scalar fields \cite{cfo-Gottlober:1989ww}--\cite{cfo-Cuzinatto:2013pva},
 which led to a two-field treatment of such models. Also, in \cite{cfo-Castellanos:2018dub}
 the correction  $R\opensquare R$ was treated as a small perturbation, and its influence on the parameters
 of cosmological perturbations was studied.

 Renormalization of the energy-momentum tensors of quantum fields in the framework of the semiclassical approach to gravity \cite{cfo-BOS1992} also leads to the inclusion of higher derivatives in modified gravity theories.

In articles \cite{cfo-Naruko:2015zze,cf-CHNM2017,cfo-Chervon:2018ths} it was shown how to reduce theory of gravity which contain in the action the Ricci scalar and its first and second derivatives to GR minimally coupled to few scalar fields. For special choices of functional dependence $f(R,(\nabla R)^2,\Box R)$ it is possible to reduce the theory to a chiral cosmological model \cite{cfo-Tsoukalas2017}. One such example and its application in cosmology is demonstrated in \cite{Chervon:2019jfu}.

In the work \cite{Chervon:2019jfu} the study of the model, using the technique described in \cite{cfo-Naruko:2015zze,cfo-Tsoukalas2017},
was continued. There were carried out a similar transition to a scalar-tensor theory by introducing Lagrangian multipliers and
 the arising auxiliary fields. Using a conformal transformation from the Jordan frame
 to the Einstein one it was obtained that the model can be represented as a two-component nonlinear
 sigma model with an interaction potential, or as a chiral cosmological model (CCM).

 In our present work we deal with gravitation in a spherically symmetric spacetime, not with cosmology. Therefore, the model  with the action
 \begin{equation}\label{act-gen}
  S_{CSGM}=\int\sqrt{-g}d^4x\left(\frac{R}{2\kappa}-
    \frac{1}{2}h_{AB}(\varphi)\varphi^A_{,\mu}\varphi^B_{,\nu}g^{\mu\nu}-W(\varphi)\right),
 \end{equation}
 (in any spacetime)
 are nothing but a self-gravitating non-linear sigma model with the potential of interaction. For the sake of brevity we will refer  to this type models as Chiral Self-Gravitating Model (CSGM) \cite{Chervon_arx_2020}. It is clear that CCM is a subset of CSGM narrowed to cosmological spaces.
Thus, in present article we study the system of gravitational and chiral fields equations of CSGM in a spherically symmetric spacetime.

The article is organised as follow.
In section \ref{CSGM} we present some brief information on the "chiral" terminology, the action of CSGM and general equations of the multiplet of kinetically and potentially interacting scalar fields in Einstein gravity.
Section \ref{EF} is devoted to description of the conformal transformations from Jordan to Einstein frame and derivation of the action for the gravity model with a kinetic curvature scalar.
The action of the resulting model in GR with two scalar fields with the potential and kinetic interaction is displayed.

In following Section \ref{SPS} the gravitational and chiral fields equations for the two component CSGM in a spherically symmetric static spacetime are presented in a general form and in harmonic coordinates.
Subsection \ref{EX} contains a simplification of the model's equations by virtue of the scaling transformation in transition from Jordan frame to Einstein one.

The section \ref{QGRS} is devoted to exact solutions
for quasi-GR model with zero potential. Two classes of one-parametric solutions are found.

In section \ref{HCAB}  we study horizons, centers and asymptotic behavior of obtained solutions, including Hawking temperature and Kretschmann scalar for a special combination of the integration constant that is particularly relevant.

Section \ref{HCAB-2} contains analysis of solutions with $k \ne 0$, including asymptotical behavior of the solution and comparison with solutions in Brans-Dicke gravity. A special case corresponding to Schwarzschild solution is studied as well.

Section \ref{ansatz} is devoted to finding, analysis and investigation of asymptotic properties of solutions with ansatz on metric components. In section \ref{EXS} we study a special case of that ansatz, corresponding to generalisation of Schwarzschild solution.
In Conclusion we summarise the obtained results and discuss new possible application of $f(R)$ gravity with higher derivatives in astrophysical experiment connected with gravitational waves.
In Appendixes {\bf A, B} and {\bf C} we present useful formulae for the deduction of the solutions and the criteria for selection of black hole and wormhole solutions.

\section{Chiral Self-Gravitating Model}\label{CSGM}

It is well known that in nuclear physics the term {\bf chiral model} was introduced for $SU(2)_L \times SU(2)_R$ model for pion field $\pi^i(x)$.
The term {\bf nonlinear sigma model} have been introduced in the work by M.~Gell-Mann and M.~Levi \cite{GellMann1960} where the chiral symmetry for fermions was considered also.
In the review by A.~Perelomov \cite{Perelomov87}  geometrical aspects of (two-dimensional, mainly) chiral models were presented
and the term "chiral model of a general type" for $n$-dimensional nonlinear manifold with Riemann metric was introduced.
Chiral models are field theory models in which the interaction is introduced not by adding the interaction Lagrangian to the free field Lagrangian, but in a purely geometric way. Namely, the Lagrangian in such models remains the same as in the case of a free fields, but constraints are imposed on the fields itself, so that now the fields $\varphi^A$ take on values already in some nonlinear manifold $M$.

Inclusion of the gravitational field in chiral models
was connected with active search of instanton and meron solutions in the case of 4-dimensional ($4D$) spacetime of Euclidian signature. First time such solutions in $4D$ model were found in the work by V.~de~Alfaro et al~\cite{deAlfaro1979}.
G.~Ivanov~\cite{Ivanov1983} independently of work \cite{deAlfaro1979} proposed "non-linear sigma model coupled to gravity" considering the Lorentz's signature metric of a spacetime and the scalar (chiral) fields as the source of gravity, besides the kinetic interaction have been introduced as the metric of a chiral (target) space. This type of models dubbed "self-gravitating sigma models".

The potential of interaction of chiral fields has been introduced by S.~Chervon in 1994~\cite{Chervon95b}.
Such model in \cite{Chervon95b} was called as "Self-gravitating nonlinear sigma model with the potential". Then in further publications in cosmological aspects, using terminology introduced by Perelomov,  the model was referred to as "chiral inflationary model" and then "chiral cosmological model".
Let us note that, the term "Chiral Cosmological Model" reflects the geometrical interactions of fields via the metric of the target (chiral) space which includes the kinetic interactions.

Summing up we can say that the term {\bf a chiral cosmological model} is introduced as a short equivalent of a self-gravitating nonlinear sigma model with the potential of interactions employed in cosmology \cite{Chervon95b,cfo-chervon2013}.

Let us note that in our present work we deal with gravitation in a spherically symmetric spacetime, not with cosmology. Therefore, the model  with the action (\ref{act-gen})
 are nothing but a self-gravitating nonlinear sigma model with the potential of interaction. As we mentioned in Introduction, for the sake of brevity we will refer to this type models as {\bf Chiral Self-Gravitating Model} (CSGM) \cite{Chervon_arx_2020}. It is clear that CCM is a subset of CSGM narrowed to cosmological spaces.

The action of CSGM
%
reads:
\begin{equation}\label{act-ccm}
  S_{CSGM}=\int\sqrt{-g}d^4x\left(\frac{R}{2\varkappa}-
    \frac{1}{2}h_{AB}(\varphi)\varphi^A_{,\mu}\varphi^B_{,\nu}g^{\mu\nu}-W(\varphi)\right),
\end{equation}
where $ R $ is the scalar curvature of the Riemannian manifold with the metric
$g_{\mu\nu}(x)$, $\varkappa=8\pi G$ is Einstein's gravitational constant and $G$ -- Newton gravitational constant, $\varphi=(\varphi^1,\ldots,\varphi^N)$~being a
multiplet of the chiral fields (we use a notation
$\varphi^A_{,\mu}=\partial_{\mu}\varphi^A =\frac{\partial
  \varphi^A}{\partial x^\mu}$),
$h_{AB}$~being the metric of the target (chiral) space with
the line element
\beq\label{tsmet}
 d s^2_{\sigma}=h_{AB}(\varphi)d\varphi^A d\varphi^B,~~A,B,\ldots =\overline{1,N}.
\eeq
The energy-momentum tensor for the model (\ref{act-ccm}) reads
\begin{equation}\label{ch-em}
  T_{\mu\nu}=h_{AB}\varphi^{A}_{,\mu}\varphi^B_{,\nu}-
  g_{\mu\nu}\left(\frac{1}{2}\varphi^A_{,\alpha}\varphi^B_{,\beta}g^{\alpha\beta}h_{AB}+
    W(\varphi)\right).
\end{equation}

The Einstein equation can be represented in the form
\beq\label{ein}
R_{\mu\nu}=\varkappa\{
h_{AB}\varphi^A_{,\mu}\varphi^B_{,\nu}+ g_{\mu \nu} W(\varphi)\},
\eeq
which simplify the derivation of gravitational dynamic equations.

Varying the action (\ref{act-ccm}) with respect to $\varphi^C$, one can
derive the dynamic equations of the chiral fields
\begin{equation}\label{4}
  \frac{1}{\sqrt{-g}}\partial_{\mu}(\sqrt{-g}h_{AB}g^{\mu\nu}\varphi_{,\nu}^A)-\frac{1}{2}\frac{\partial
    h_{BC}}{\partial\varphi^A}\varphi^C_{,\mu}\varphi^{B}_{,\nu}g^{\mu\nu}-W_{,A}=0,
\end{equation}
where $W_{,A}=\frac{\partial W}{\partial\varphi^A}$.
Considering the action (\ref{act-ccm}) in the framework of
cosmological spaces, we arrive to a chiral cosmological model
\cite{Chervon95b}, \cite{Chervon95a}, \cite{cfo-chervon2013}, \cite{cfo-CHFK2015}, \cite{ch00mg}, \cite{ch02gc}.

In the present work we will deal with the $2$-component diagonal chiral metric
$$
ds^2=h_{11}d\chi^2 + h_{22}(\chi,\phi)d\phi^2, ~~h_{11}=const.
$$
For the sake of completeness we display the chiral fields equation (\ref{4}) for this case
\begin{equation}
h_{11}\ddot{\chi}+3h_{11}H\dot{\chi}-\frac{1}{2}\frac{\partial h_{22}(\chi,\phi)}{\partial \chi}\dot{\phi}^2+\frac{\partial W(\chi,\phi)}{\partial \chi}=0,
\end{equation}

\begin{equation}
3 h_{22}(\chi,\phi)H\dot{\phi}+\frac{d}{dt}(h_{22}\dot{\phi})- \frac{1}{2}\frac{\partial h_{22}(\chi,\phi)}{\partial \phi}\dot{\phi}^2+\frac{\partial W(\chi,\phi)}{\partial \phi}=0.
\end{equation}

In the next section we will obtain concrete forms for $h_{22}(\chi,\phi)$ and $W(\chi,\phi) $.

\section{The $f(R)$ gravity model with a kinetic scalar curvature in Einstein frame}\label{EF}

In the work \cite{cfo-Naruko:2015zze} the authors consider the most general form of $f(R)$ gravity with higher derivatives of first and second order with respect to Ricci scalar.
 Such a theory of gravity is described by the action
\beq
\label{cfo-Gen}
	S_{\rm gen} = \int d^{4}x\sqrt{-g_S}\, f \left(R_S,(\nabla R_S)^{2},\square R_S \right),
\eeq
where $(\nabla R)^{2}=g^{\mu\nu}\nabla_{\mu}R\nabla_{\nu}R$ and the index ${}_S$ corresponds to a choice of the initial metric.

Let us suppose that the action (\ref{cfo-Gen}) is written in Einstein's frame (E-frame) and we study the truncated model with the action
\begin{eqnarray}\label{cfo-bfR}
\nonumber
S_{fRR'}=\int d^{4}x\sqrt{-g}\, f\left(R,(\nabla R)^{2} \right), \\
f(R,(\nabla R)^2)=f_1(R)+X(R)\nabla_\mu R\nabla^\mu R.
\end{eqnarray}
%
Here we omitted the index ${}_S$. Also in (\ref{cfo-bfR}) and hereafter $f_1(R)$ and $X(R)$ are $\textsl{C}^1$ functions  of a scalar curvature.
Let us note that setting $f_1(R)=R$ and $X(R)=0$ we reduce the model to GR.
The model (\ref{cfo-bfR})
have been studied for cosmological applications in \cite{cf-CHNM2017}, \cite{cfo-Chervon:2018ths}, \cite{Chervon:2019jfu}.

Following by prescription of \cite{cfo-Naruko:2015zze} and using the set of Lagrangian multipliers (and related auxiliary fields) one can bring the action to the form
\begin{equation}\label{lambdaR}
S_{JfRR'}=\int d^{4}x\sqrt{-g_J}\,\left[ f\left(\phi,(\nabla \phi)^{2}\right)+ \lambda_L(R_J-\phi)\right],
\end{equation}
were $\lambda_L(x)$ is one of dynamical fields. Moreover, under such procedure we have formal replacements of $R$ by $\phi$ and $\nabla_\mu R$ by $\nabla_\mu \phi$.
As one can see from (\ref{lambdaR}), we arrive to the theory with non-minimal interaction gravity with the scalar field $\lambda_L$. So we will consider the model (\ref{lambdaR}) as gravity in Jordan frame (J-frame) and have possibility to move to the E-frame via a conformal transformation \cite{Chervon:2019jfu}
\begin{equation}
g_{\mu \nu}^E = \Omega^2 g_{\mu \nu}^J,~~\Omega^2 = 2 \lambda_L.
\end{equation}
 After that we obtain Einstein gravity with two noncanonical scalar fields $\lambda_L$ and $\phi$ with kinetic terms and the potential. To reduce a kinetic part of $\lambda_L$ to a canonical form one can introduce the new scalar field $\chi$ by the following way
\begin{equation}
\lambda_L=\exp\left(\sqrt{(2/3)} \chi \right).
\end{equation}

Finally the model (\ref{cfo-bfR}) is transformed to
the Einstein scalar fields gravity \cite{cfo-Naruko:2015zze},\cite{Chervon:2019jfu} with the action
\begin{eqnarray}\nonumber
 S_{EfRR'}=\int d^4x \sqrt{-g_E}\bigg(\frac{R_E}{2\varkappa} - \frac{1}{2}g^{\mu\nu}_E\chi_{,\mu}\chi_{,\nu}+\\
  \frac{1}{4}f_1(\phi)e^{-2\sqrt{2/3}\chi}-
 \frac{1}{4}\phi e^{-\sqrt{2/3}\chi}
 \label{act-2}
  + \frac{1}{2}Xe^{-\sqrt{2/3}\chi} g^{\mu\nu}_E\phi_{,\mu}\phi_{,\nu} \bigg).
\end{eqnarray}
where subscript ${}_E$ denotes the Einstein's frame.


The second scalar field $\phi$ takes the nonlinearity of $f_1(R)$ and in some sense can be in correspondence with dependence $f_1(\phi)$ \cite{cfo-Naruko:2015zze},\cite{Chervon:2019jfu}.
Details of the derivation of the action (\ref{act-2}) can be found in \cite{Chervon:2019jfu}.

The action $S_{EfRR'}$ (\ref{act-2}) can be presented in the form of a chiral cosmological model \cite{cfo-chervon2013} with two chiral fields
 $\varphi^1=\chi$, $\varphi^2=\phi$, and 2D metric of the target space
\begin{equation}\label{ch-ds2}
	ds^2 = d\chi^2 - e^{-\sqrt{2/3}\chi}X(\phi )d \phi^2,~~ h_{11}=1,~~h_{22}=- e^{-\sqrt{2/3}\chi}X(\phi ).
\end{equation}

From (\ref{act-2}) one can easily to extract the interaction potential
\begin{equation}
\label{W}
W=\frac{1}{4}e^{-\sqrt{2/3}\chi}\left(\phi-e^{-\sqrt{2/3}\chi}f_1(\phi)\right).
\end{equation}
%


Further we consider CSGM with fixed forms of chiral metric (\ref{ch-ds2}) and the potential (\ref{W}).

\section{Model equations for spherically symmetric static spacetime}\label{SPS}

We start from standard representation of the spherically symmetric spacetime \cite{Bronnikov} in the following diagonal form:

\begin{equation}\label{SphSy_mbr}
  ds^2=-e^{2\nu (u)}dt^2+e^{2\lambda (u)} du^2 +e^{2\beta (u)} \left(d\theta^2+\sin^2 \theta d\varphi^2 \right).
\end{equation}
Einstein equations (\ref{ein}) can be derived based on components of Ricci tensor and they are
\begin{equation}\label{E-0br}
\exp \left[-2\lambda+2 \nu\right] \left( \nu''  +(\nu')^2 +
\nu' (2\beta' - \lambda' ) \right)=-\varkappa e^{2\nu}W,
\end{equation}
%
$$
-2 \beta''-\nu''+\lambda' (\nu'+2\beta')-(\nu')^2 -2(\beta')^2
$$
%
\begin{equation}\label{E-1br}
=\varkappa \left(h_{11}(\chi')^2+h_{22}(\phi')^2+e^{2\lambda} W\right),
\end{equation}

\begin{equation}\label{E-2br}
1+ \exp \left[ -2\lambda+2 \beta\right] \left(- \beta''  -2 (\beta')^2 +
\beta' (-\nu' + \lambda' ) \right)=\varkappa e^{2\beta}W,
\end{equation}
where $W$ is defined in (\ref{W}).

Equations for chiral fields are
\begin{eqnarray}
h_{11}\chi'' + h_{11}(\nu'-\lambda'+2\beta')\chi'-\frac{1}{2}\frac{\partial h_{22}}{\partial \chi}(\phi')^2=e^{2\nu}\frac{\partial W}{\partial \chi},\\
(h_{22} \phi')' + h_{22}(\lambda'-\nu'+2\beta')\phi'=e^{2\lambda}\frac{\partial W}{\partial \phi}.
\end{eqnarray}

Substituting chiral metric components $h_{11}, h_{22}$ from (\ref{ch-ds2}) and the potential $W$ from (\ref{W}) one can obtain
 \begin{eqnarray}\label{fleq-1}
 \nonumber
h_{11}\chi'' + h_{11}(\nu'-\lambda'+2\beta')\chi'-
\frac{1}{\sqrt{6}}e^{-\sqrt{\frac{2}{3}}\chi}X(\phi)(\phi')^2\\
+ e^{2\lambda}\frac{1}{\sqrt{6}}\left[ \frac{1}{2}e^{-\sqrt{\frac{2}{3}}\chi} \phi+e^{-2\sqrt{\frac{2}{3}}\chi}f_1(\phi)\right]=0,
\end{eqnarray}
\begin{eqnarray}\label{fleq-2}
 \nonumber
X(\phi) \phi'' + (\nu'-\lambda'+2\beta')X(\phi)\phi'-\sqrt{\frac{2}{3}}X(\phi)\chi'\phi' +\frac{1}{2} X_{,\phi}(\phi')^2\\
+\frac{1}{4}e^{2\lambda}\left(1-e^{-\sqrt{\frac{2}{3}}\chi}f_{1,\phi}\right)=0,
\end{eqnarray}
where prime denotes to the derivative of variable $u$, $X_{,\phi}=\frac{dX(\phi)}{d\phi}$, and $f_{1,\phi}=\frac{df_1(\phi)}{d\phi} $.

\subsection{Equations in harmonic coordinates}

Let us choose the harmonic coordinates where we have the connection between exponents of metric functions of the form
\begin{equation}\label{hg}
\lambda=2\beta+\nu.
\end{equation}

Einstein equations (\ref{E-0br})-(\ref{E-2br}) take the following form:
\begin{equation}\label{Eh-0br}
e^{-4\beta}\nu'' =-\varkappa e^{2\nu}W,
\end{equation}

\begin{equation}\label{Eh-1br}\nonumber
-2\beta''-\nu''+2(\beta')^2 +4\beta'\nu'=
\end{equation}
\begin{equation}
=\varkappa
\left(h_{11}(\chi')^2+h_{22}(\phi')^2+e^{4\beta+2\nu} W\right),
\end{equation}

\begin{equation}\label{Eh-2br}
1- \beta ''\exp \left[-2 \beta-2\nu\right]   =\varkappa e^{2\beta}W.
\end{equation}

Chiral fields equations (\ref{fleq-1})-(\ref{fleq-2}) for harmonic gauge (\ref{hg}) reduce to
\begin{eqnarray}
\nonumber
h_{11}\chi'' -\frac{1}{\sqrt{6}} e^{-\sqrt{\frac{2}{3}\chi}}X(\phi)(\phi')^2\\
-e^{4\beta+2\nu}\left[-\frac{1}{2\sqrt{6}} e^{-\sqrt{\frac{2}{3}\chi}}\phi+\frac{1}{\sqrt{6}} e^{-2\sqrt{\frac{2}{3}\chi}}f_1(\phi) \right]=0,
\end{eqnarray}
\begin{equation}
-\sqrt{\frac{2}{3}}X(\phi)\phi'\chi'+\frac{1}{2}X_{,\phi}(\phi')^2+
X(\phi)\phi''
+\frac{1}{4}e^{4\beta+2\nu} \left(1-e^{-\sqrt{\frac{2}{3}\chi}}f_{1,\phi}\right)=0.
\end{equation}
%

From equations (\ref{Eh-0br}) and (\ref{Eh-2br}) one can obtain the following relation for the metric components:
\begin{equation}\label{b-nu}
\beta''-\nu''=e^{2\beta+2\nu}.
\end{equation}

Now, we consider a some classes of exact solutions of equations (\ref{Eh-0br})--(\ref{b-nu}) based on a special choice of model parameters.

\subsection{Special case of scaling transformation}\label{EX}

Following by special suggestion in cosmology \cite{cfo-Tsoukalas2017,Chervon:2019jfu}  let us study the case when the scalar field  $\chi$ is equal to special constant $\chi= -\sqrt{\frac{3}{2}}\ln 2$. This value of $\chi$ corresponds to identical conformal transformation with $\Omega^2=1$.

The gravitational and chiral field equations will take the following form
\begin{equation}\label{30-nu}
e^{-(4\beta+2\nu)}\nu''=-\varkappa \left(\phi/2-f_1(\phi)\right),
\end{equation}

\begin{equation}\label{31-X}
-2\beta''-\nu''+2(\beta')^2 +4\beta'\nu'=\varkappa \left(-2X(\phi)(\phi')^2+e^{4\beta+2\nu}(\phi/2-f_1(\phi))\right),
\end{equation}

\begin{equation}\label{32-beta}
1-\beta'' e^{-2\beta-2\nu}=\varkappa e^{2\beta}(\phi/2-f_1(\phi)),
\end{equation}

\begin{equation}\label{f-2-spec}
-2X(\phi) (\phi')^2-e^{4\beta+2\nu}(-\phi+4f_1(\phi))=0.
\end{equation}

\begin{equation}\label{34-phi}
2X(\phi)\phi''+X_{,\phi} (\phi')^2+e^{4\beta+2\nu}(1/2-f_{1,\phi})=0.
\end{equation}

The equation (\ref{f-2-spec}) may be considered as additional one, since it does not appear in the absence of $\chi$ (no variation on $\chi$).

Further investigation in the article concern only the special case when $ \chi= -\sqrt{\frac{3}{2}}\ln 2$.

\section{Quasi-GR solutions. Case $f_1(\phi)=\phi/2$}\label{QGRS}

Setting $\chi= -\sqrt{\frac{3}{2}}\ln 2$ and $f_1(\phi)=\phi/2$ one can find that the potential vanish: $W=0$.
Such suggestion reduce the model to GR, if we additionally set $X(\phi)=0$ in the action (\ref{act-2}).

For the case $f_1(\phi)=\phi/2$ we have the system of equations with the solution of $\nu(u)$
\begin{equation}\label{nu-1G}
\nu''=0,~~\nu=A_1u+A_2,~~A_1, A_2 - const.,
\end{equation}

\begin{equation}\label{35G}
-2\beta''+2(\beta')^2 +4\beta'A_1=\varkappa \left(-2X(\phi)(\phi')^2\right),
\end{equation}

\begin{equation}\label{sol-betaG}
1-\beta'' e^{-2\beta-2A_1u-2A_2}=0,
\end{equation}

\begin{equation}\label{37G}
2X(\phi)\phi''+X_{,\phi}(\phi')^2=0,
\end{equation}

We can rewrite the equation (\ref{37G}) as follows
\begin{equation}\label{G1}
2X(u)\phi''+X' \phi'=0.
\end{equation}
The solution is
\begin{equation}\label{G2}
X(u)(\phi')^{2}=X(\phi)(\phi')^{2}=C,
\end{equation}
where $C$ is the constant of integration.
Thus, one can substitute the solution (\ref{G2}) into equation (\ref{35G}) to find the function $\beta(u)$.

The equation (\ref{G2}) gives for us an interesting possibility to find appropriate theory parameter $X(\phi)$ when the functional dependence $\phi$ on $u$ is given. Some examples can be found in \cite{Chervon_arx_2020}.

\subsection{Two classes of exact solutions}\label{ex_sols}

One can write following exact solutions of equations (\ref{35G})--(\ref{sol-betaG}) in terms of the constant $k\equiv A^{2}_{1}-\varkappa C$:
\begin{itemize}

\item The first class of exact solutions, corresponding to the case $k=0$, is represented by 
\begin{equation}\label{G5}
\beta_{1}(u)=-A_{1}u-A_{2}-\ln|u-u_{\ast}|.
\end{equation}

For this solution, the connection with GR is defined by the condition $A_{1}=0 \Rightarrow C=0$. If $C=0$ in expression (\ref{G2}) we have two possibilities $X=0$ or $\phi = const.$ Both cases leads to vacuum (degenerate) case of GR. Definitely, if we set $X(\phi)=0$ any scalar field $\phi$ will satisfy field equation (\ref{G2}).

\item The second class of exact solutions, corresponding to the case $k\neq0$, is represented by

\begin{equation}\label{G3}
\beta_{2}(u)=-A_{1}u-A_{2}-\ln\left[\frac{c}{2}e^{\pm\sqrt{k}(u-u_{\ast})}-\frac{1}{2kc}e^{\mp\sqrt{k}(u-u_{\ast})}\right],
\end{equation}
where $c\neq 0$ and $u_{\ast}$ are some constants.


If $k=A^{2}_{1}$ ($C=0$) in expression (\ref{G2}) once again we have two possibilities $X=0$ or $\phi = const.$ So the connection with GR is defined by the condition $k=A^{2}_{1}$.
\end{itemize}

Thus, in this case, one can write the components of the metric (\ref{SphSy_mbr}) as
\begin{eqnarray}
\label{M1}
&&e^{2\nu} =e^{2(A_1u+A_2)},\\
\label{M2}
&&e^{2\beta} = \Theta(u)e^{-2\nu},\\
\label{M3}
&&e^{2\lambda} = e^{2(2\beta+\nu)}=\Theta^{2}(u)e^{-2\nu},
\end{eqnarray}
where function $\Theta(u)$ is defined as follows
$$
\Theta(u) =\left\{
  \begin{array}
  [c]{ll}%
   \Theta_{1}(u)\equiv(u-u_{\ast})^{-2},~~~~~~~~~~~~~~~~~~~~~~~~~~~~~~~\mbox{for first class}, \\
  \Theta_{2}(u)\equiv\left[\frac{c}{2}e^{\pm\sqrt{k}(u-u_{\ast})}-\frac{1}{2kc}e^{\mp\sqrt{k}(u-u_{\ast})}\right]^{-2},~~~
  \mbox{for second class},
   \\
\end{array}
\right.
$$

We will analyse the solutions for real functions $\nu(u)$, $\lambda(u)$ and $\beta(u)$ in metric \eref{SphSy_mbr} and, respectively, one has the condition $\Theta_{2}(u)>0$.

Now, we consider particular solutions for $\Theta_{2}(u)$ under this conditions.
For the case $k>0$ and $c=\pm\frac{1}{\sqrt{k}}$ one has
\begin{equation}
\label{ThetaA}
\Theta^{(A)}_{2}(u)=k\left(\sinh\left[\sqrt{k}(u-u_{\ast})\right]\right)^{-2}.
\end{equation}

For the case $k<0$ and $c=\pm\frac{1}{\sqrt{-k}}$ one has
\begin{equation}
\label{ThetaB}
\Theta^{(B)}_{2}(u)=|k|\left(\cos\left[\sqrt{|k|}(u-u_{\ast})\right]\right)^{-2}.
\end{equation}

Therefore, one can generate the exact solutions for chosen dependencies $X=X(\phi)$ (or $\phi=\phi(u)$) taking into account the condition (\ref{G2}) and reconstruct the function $\phi=\phi(u)$ (or $X(\phi)$) from this parametric representation.

\section{Horizons, centers and asymptotic behavior for the solution with $k=0$}\label{HCAB}

Let us study the case $C=\frac{A^{2}_{1}}{\varkappa}$ or $\Theta_{1}(u)$ in more details.
%
%
The metric components of the solution for this case are described by the following formulae
\begin{eqnarray}\label{beta-44}
e^{2\nu} &=& e^{2(A_1 u+A_2)},\\
e^{2\lambda} &=& \frac{e^{-  2(A_1 u+A_2)}}{(u - u_*)^4},\\
\label{e-beta-49}
e^{2\beta} &=& \frac{e^{- 2(A_1 u +A_2)}}{(u - u_*)^2}.
\end{eqnarray}

In our analysis of possible horizons, centers and asymptotic behavior one can use  criteria for selection of black holes and wormholes solution reproduced from \cite{Bronnikov97,kb-96-GC} in \ref{CRIT}. One can find in \ref{HKRET} also formulae for calculation of Hawking temperature and Kretschmann scalar in harmonic coordinates.

\subsection{Properties of the solution with $k \equiv A_1^2-\varkappa C=0$}

We displayed the solution in beginning of this section in formulae (\ref{beta-44})-(\ref{e-beta-49}). Now we study
some special cases of this solution.

\subsubsection{$A_1 > 0$}

\subsubsection*{Horizon}
There is a possible horizon when $ e^{2\nu}\rightarrow 0 $, what is possible if $u \rightarrow - \infty$.

For this asymptote we have
\begin{itemize}
\item $u \rightarrow - \infty$:
\begin{eqnarray}
e^{2\nu} &\rightarrow& 0,\\
e^{2\lambda} &\rightarrow& \infty,\\
e^{2\beta} &\rightarrow& \infty.
\end{eqnarray}

\end{itemize}

Direct checking of the Kretschmann scalar gives $\mathscr{K} \rightarrow 0.$ The Hawking temperature $T_H \rightarrow 0 $.
\begin{table}
\begin{center}
\begin{tabular}{ |c|c|c|c|c|c|c|c|c| }
 \hline
 $e^{2\nu}$ &$e^{2\beta}$ &$e^{2\lambda}$ & $u$ & ${\bf C1}$ & ${\bf C2}$& ${\bf C3}$& ${\bf C4}$& ${\bf C5}$ \\
  \hline
 $0$& $\infty$ & $\infty$ & $-\infty$ &Yes & No &Yes & Yes &Yes \\
  \hline
%
   \end{tabular}
\end{center}
\caption{\label{tab1-1} The correspondence to horizon criteria for solutions \eref{beta-44}-\eref{e-beta-49}.}
\end{table}

The criteria for the solution \eref{beta-44}-\eref{e-beta-49} (when $A_1>0)$ are presented in the Table \ref{tab1-1}.
Thus we obtained cold black hole solution or, by terminology of \cite{Bronnikov97}, type B black hole.

\subsubsection*{Center}
A center occurs when $r\equiv e^\beta=0$. To get such result we must request
\begin{itemize}
\item $u \rightarrow \infty$
\end{itemize}
Under this condition we have $e^{2\beta} \rightarrow 0$ and
 $\mathscr{K} \rightarrow \infty $.
For $A_1 > 0$, the center occurs when $u \rightarrow \infty$. This is in the range $ u_* <  u$ corresponding, with respect to the previous determined black hole solution, to the change $\sqrt{k} \rightarrow - \sqrt{k}$, that is, a naked singularity, as expected.

\subsubsection*{Asymptotical flatness}
Let us check asymptotical flatness of the solution. For this target we must consider remote distances when $r\equiv e^\beta \rightarrow \infty $.
This relation we can obtain
on the other asymptote,

\begin{itemize}
\item $u \rightarrow u_*$.
\end{itemize}

For further investigation, we use the variable $x$ defined from the relation $u = x + u_*$. Then (tending $x$ to $0$) the solution can be approximatively described by
the metric,
\begin{eqnarray}
ds^2 = -\frac{dt^2}{C^2} + C^2\frac{dx^2}{x^4} + \frac{C^2}{x^2}d\Omega^2,
\end{eqnarray}
where $C = e^{-2A_1u_*-A_2}$ is a constant. The time coordinate can be redefined in order to absorb this constant.
Defining the new coordinate,
\begin{eqnarray}
y = \frac{C}{x},
\end{eqnarray}
the metric takes the form,
\begin{eqnarray}
ds^2 = - d\bar t^2 + dy^2  + y^2d\Omega^2.
\end{eqnarray}
That is we get a Minkowski spacetime. Hence the solution is asymptotically flat.

\subsubsection{$A_1 < 0$}

The configuration is similar to the previous case.
The horizon is situated at $u \rightarrow \infty$ and the spatial (Minkowski) infinity at $u = u_*$. In more detail:

\subsubsection*{Horizon}

There is a possible horizon when $ e^{2\nu}\rightarrow 0 $, what is possible if $u \rightarrow  \infty$.

For this asymptote we have
\begin{itemize}
\item $u \rightarrow \infty $:

\begin{eqnarray}
e^{2\nu} &\rightarrow& 0,\\
e^{2\lambda} &\rightarrow& \infty,\\
e^{2\beta} &\rightarrow& \infty.
\end{eqnarray}

\end{itemize}

Direct checking of the Kretschmann scalar gives $\mathscr{K} \rightarrow 0.$
The Hawking temperature $T_H \rightarrow 0 $.

\begin{table}
\begin{center}
\begin{tabular}{ |c|c|c|c|c|c|c|c|c| }
 \hline
 $e^{2\nu}$ &$e^{2\beta}$ &$e^{2\lambda}$ & $u$ & ${\bf C1}$ & ${\bf C2}$& ${\bf C3}$& ${\bf C4}$& ${\bf C5}$ \\
  \hline
 $0$& $\infty$ & $\infty$ & $\infty$ &Yes & No &Yes & Yes &Yes \\
  \hline
%
   \end{tabular}
\end{center}
\caption{\label{tab1-2} The correspondence to horizon criteria for solutions \eref{beta-44}-\eref{e-beta-49}.}
\end{table}

The result for the solution \eref{beta-44}-\eref{e-beta-49} (when $A_1<0)$ is represented in the Table \ref{tab1-2}. It is clear that we obtained cold black hole solution.

\subsubsection*{Center}
The center occurs when $r\equiv e^\beta=0$. To get such result we must request
\begin{itemize}
\item $u \rightarrow -\infty$
\end{itemize}

Under this condition we have $e^{2\beta} \rightarrow 0$ and $\mathscr{K} \rightarrow \infty $.
Let us note that
$u \rightarrow - \infty$ is, for $A_1 < 0$, in the domain $ - \infty < u < u_*$ which describes a naked singularity.

\subsubsection*{Asymptotical flatness}

Let us study asymptotical flatness of the solution.
For this we must consider remote distances when $r\equiv e^\beta \rightarrow \infty $.
This relation we can obtain on the other asymptote,

\begin{itemize}
\item $u \rightarrow u_*$
\end{itemize}

We may perform a similar analysis as for the case $A_1>0$
using the new variable $x$ defined from the relation $u = x + u_*$.
We get once again a Minkowski spacetime. Hence the solution is asymptotically flat at $u=u_*$.

\subsubsection{Searching for wormhole solution}

It is clear that one can choose two value of generalised radial coordinate $u$ as two necessary surfaces: $u=u_{-\infty}=-\infty$ and $u=u_*$. Then the criteria {B1, B2, B3} are true, but {B4} is not true. Thus, there is no wormhole solution for this case.


\section{Analysis of solutions with $k \ne 0$ and $\Theta_{2}(u)>0$}\label{HCAB-2}

In the case when $C\neq\frac{A^{2}_{1}}{\varkappa}$
we will consider the solutions for $\Theta_{2}(u)>0$.
Let us analyse the cases with special relations between constants and with initially chosen the following condition for an arbitrary constant $u_{\ast}=0$, that does not change the generality of analysis.

\subsection{Solution $\Theta^{(A)}_{2}(u)$ for $k>0$}

The solution is described by
\begin{eqnarray}
e^ {2\nu} &=& e^{2 (A_1 u + A_2)},\\
e^ {2\beta} &=& \Theta(u)e^ {-2\nu},\\
e^{2\lambda} &=& \Theta(u)^2 e^{-2\nu},\\
\Theta(u) &=& \frac{k}{\sinh^{2}\left(\sqrt{k}u\right)}.
\end{eqnarray}

We will study only the case $A_1>0$ since the $A_1<0$ is just the reversal configuration, with the same properties.

\begin{itemize}
\item $A_1 > 0$.
\end{itemize}

In this case, the possible horizon is located at $u \rightarrow - \infty$. For the coefficient
of the angular part, $e^ {2\beta}$, we have,
\begin{eqnarray}
e^ {2\beta} \rightarrow e^{2(\sqrt{k} - A_1)u}.
\end{eqnarray}
Hence we have three possibilities:
\begin{enumerate}
\item $A_1 - \sqrt{k} > 0 $, and the area of the horizon is infinite, indicating a cold black hole;
\item $A_1 - \sqrt{k} = 0$, and the area of the horizon is finite, indicating a Schwarzschild-type horizon;
\item $A_1 - \sqrt{k} < 0$, and the area of the horizon goes to zero, indicating a singularity, a center.
\end{enumerate}
For the radial metric function, $e^ {2\lambda}$, we have,
\begin{enumerate}
\item $A_1 - 2\sqrt{k} > 0 $, and the function diverges, as it happens with the Schwarzschild black hole;
\item $A_1 - 2\sqrt{k} = 0$, and the function is finite;
\item $A_1 - 2\sqrt{k} < 0$, and the radial metric function goes to zero, indicating a singularity, a degenerate metric.
\end{enumerate}

We may characterise the other asymptotic by the zero of the $\sinh$ function. It happens for
$u = 0$.
In this case, we have a minkowskian asymptotical space-time.
In brief, it seems we have necessary cold black holes (infinite horizon surface) only for $A_1 > 2\sqrt{k}$.

\subsection{Solution $\Theta^{(B)}_{2}(u)$ for $k<0$}
The solution is described by
\begin{eqnarray}
e^ {2\nu} &=& e^{2 (A_1 u + A_2)},\\
e^ {2\beta} &=& \Theta(u)e^ {-2\nu},\\
e^{2\lambda} &=& \Theta(u)^2 e^{-2\nu},\\
\Theta(u) &=& \frac{|k|}{\cos^2(\sqrt{|k|}u)}.
\end{eqnarray}

Once again we will study only the case $A_1>0$ since the $A_1<0$ is just the reversal configuration, with the same properties.

\begin{itemize}
\item $A_1 > 0$.
\end{itemize}

The coordinate $u$ is defined in the range $- \frac{\pi}{2} \leq \sqrt{|k|}u \leq \frac{\pi}{2}$.
Making the approximation in the two asymptotical regions,
\begin{eqnarray}
\sqrt{|k|}u = \pm \frac{\pi}{2} \mp y, \quad y \sim 0,
\end{eqnarray}
we end up with the same minkowskian metric,
\begin{eqnarray}
ds^2 = d\bar t^2 - dr^2 - r^2d\Omega^2,
\end{eqnarray}
where,
\begin{eqnarray}
\bar t &=& e^{2\nu_\pm}t,\\
r &=& e^{-2\nu_\pm}y,
\end{eqnarray}
with $e^{2\nu_\pm} = e^{\pm \frac{A_1}{\sqrt{|k|}}\pi + A_2}$.

Hence this case corresponds to a wormhole with two minkowskian asymptotic.

\subsection{Comparison with solution in Brans-Dicke gravity}

The metric for a large class of solutions found previously has the form,
\begin{equation}
\label{metric1}
ds^2 = - e^{-2bu}dt^2 + \frac{e^{2bu}}{s^2(k,u)}\biggr\{\frac{du^2}{s^2(k,u)} + d\Omega^2\biggl\},~~d\Omega^2=d\theta^2+ \sin^2 \theta d\varphi^2,~~b=const,
\end{equation}
where
$$
s(k,u) =\left\{
  \begin{array}
  [c]{ll}%
  \frac{\sinh\sqrt{k}u}{\sqrt{k}},~~~~k > 0,\\
  u,~~~~~~~~~~~k = 0,\\
  \frac{\sin\sqrt{-k}u}{\sqrt{-k}},~~~k < 0.
  \end{array}
\right.
$$

In Ref. \cite{kb1}, similar solutions were found in the context of the Brans-Dicke theory. However, there is a conformal factor due to the non-minimal coupling and the metric in the Brans-Dicke case takes the form,
\begin{eqnarray}
ds^2 = \phi^{-1}\biggr\{-e^{-2bu}dt^2 + \frac{ e^{2bu}}{s^2(k,u)}\biggr[\frac{du^2}{s^2(k,u)} + d\Omega^2\biggl]\biggl\},
\end{eqnarray}
 with
 \begin{eqnarray}
 \phi = e^{Cu/\sqrt{|\omega + 3/2|}},
 \end{eqnarray}
 where $C$ is a constant which plays the r\^ole of a scalar charge associated with the scalar field.
 These solutions contain black holes and wormholes, except for the case $k < 0$ where only wormhole solutions are possible.

In opposition to our case represented by metric (\ref{metric1}), in the Brans-Dicke case there is a restriction in the parameters given by,
\begin{eqnarray}
2k\mbox{sign}k = 2b^2 + \epsilon C^2,
\end{eqnarray}
where $C$ is the charge associated with the scalar field, and $\epsilon = \pm 1$ indicates if the scalar field is phantom (sign minus) or normal (sign plus).
The notion of phantom field for our case is less obvious than in the Brans-Dicke case.

In Ref. \cite{kb1} the conditions to extend the metric beyond the horizon have been established. This is a crucial condition to have black hole solutions. We can follow essentially the same steps for the metric (\ref{metric1}).

Let us consider the metric,
\begin{eqnarray}
ds^2 = - e^{-2bu}dt^2 + \frac{e^{2bu}}{\sinh^4(\sqrt{k}u)}k^2du^2
+  \frac{e^{2bu}}{\sinh^2(\sqrt{k}u)}kd\Omega^2.
\end{eqnarray}

We write,
\begin{eqnarray}
e^{-2\sqrt{k}u} = 1 - \frac{2\sqrt{k}}{r} = P.
\end{eqnarray}

With this definition, the metric becomes,
\begin{eqnarray}
\label{metric2}
ds^2 = -P^adt^2 + P^{-a}du^2
+  P^{1 - a}r^2d\Omega^2,
\end{eqnarray}
with
\begin{eqnarray}
a = \frac{b}{2\sqrt{k}}.
\end{eqnarray}

The metric is extensible if
\begin{eqnarray}
\label{cond1}
a = l + 1, \quad l \in {\cal N_+}.
\end{eqnarray}

We can compare directly with the results for the Brans-Dicke case. In Ref. \cite{kb1} the metric, after similar transformations, is written
as,
\begin{eqnarray}
ds^2 = P^{-\xi}\biggr(- P^adt^2 + P^{-a}du^2
+  P^{1 - a}r^2d\Omega^2).
\end{eqnarray}
The constraint in the parameters read now,
\begin{eqnarray}
\label{res}
(3 + 2\omega)\xi = 1 - a.
\end{eqnarray}
The solution can be extended beyond the horizon if,
\begin{eqnarray}
a = \frac{l + 1}{l - n},\\
\xi = \frac{l - n - 1}{l - n},
\end{eqnarray}
such that $l - 2 \geq n \geq 0$.
Hence, $\xi = 0$ implies $a = 1$, that is, $l = 0$, $n = - 1$.

Our case is characterised by the absence of conformal factor and also by the absence of the constraint (\ref{res}). This means $\xi =  0$ implying $l = n + 1$. The result  (\ref{cond1}) comes out directly after redefining $l + 1 \rightarrow l$ and express the situations for which the notion of black hole can be recovered including the analyticity of the metric in crossing the horizon.
For the metric (\ref{metric2}) with (\ref{cond1}), the Schwarzschild solution implies $l = 0$.
The metric (\ref{metric1}) represents a black with a causal structure similar to the Schwarzschild black hole for $a$ odd ($l$ even) and similar to the extremal Reissner-Nordstr\"om black hole for $a$ even ($l$ odd).

\section{Ansatz $\beta=m \nu $}\label{ansatz}

Choosing the relation between the exponents of metric functions
\begin{equation}\label{anz1}
\beta = m \nu,~~ m=const.,~~m \ne 1,~~m \ne -1,
\end{equation}
we can solve the equation \eref{b-nu} which takes the following form
\begin{equation}\label{dd-nu}
\nu''=\frac{1}{m-1}e^{2\nu (m+1)}.
\end{equation}

With the ansatz \eref{anz1} and choosing the $f_1(\phi)$ as
\begin{equation}\label{f2-def}
f_1(\phi)=\phi/2+K_2 f_2(\phi),~~K_2=const.
\end{equation}
the gravitational and field equations take the following form
\begin{equation}\label{30-nu-az}
e^{-2(2m+1)\nu}\nu''=\varkappa K_2 f_2(\phi),
\end{equation}

\begin{equation}\label{31-X-az}
-(2m+1)\nu''+2m(m+2)(\nu')^2=\varkappa \left(-2X(\phi)(\phi')^2-e^{2(2m+1)\nu} K_2 f_2(\phi)\right),
\end{equation}

\begin{equation}\label{32-beta-az}
1-m\nu'' e^{-2(m+1)\nu}=-\varkappa e^{2m\nu}K_2 f_2(\phi),
\end{equation}

\begin{equation}\label{34-phi-az}
2X(\phi)\phi''+X_{,\phi} (\phi')^2-e^{2(2m+1)\nu}K_2f_{2,\phi}=0.
\end{equation}

Taking into account \eref{dd-nu}, equation \eref{30-nu-az} leads to
\begin{equation}\label{f-2-nu}
f_2(\phi)=\frac{1}{\varkappa K_2 (m-1)}e^{-2m\nu}.
\end{equation}

Substitution \eref{f-2-nu} into \eref{32-beta-az} gives the identity.
Also we can find the dependence $f_2=f_2(\phi(u))=f_2(u)$ from \eref{f-2-nu} using the solutions of equation \eref{dd-nu}.

We can obtain the general form for the combination $X(\phi)(\phi')^2$ if we insert $\nu''$ from \eref{dd-nu}
and the first integral of \eref{dd-nu}
\begin{equation}\label{nu-m1}
\nu'^2=\frac{1}{m^2-1} e^{2(m+1)\nu}+C_1
\end{equation}
into equation \eref{31-X-az}. As the result we get
\begin{equation}\label{Xphi}
\left[\frac{2m}{m^2-1}\right] e^{2(m+1)\nu}  + 2m(m+2)C_1=-2\varkappa X(\phi)(\phi')^2.
\end{equation}

Let us note that
we can not obtain the constant combination $X(\phi)(\phi')^2=const.$  Thus the case considered in previous section  is not compatible with ansatz approach.

\subsection{The solution $C_1=0$}\label{ss8-1}

If we choose $C_1=0$ then we get from (\ref{nu-m1}) $m^2>1$.
We may consider two branches of solution in this case. Namely
\begin{equation}\label{C1=0a}
\nu_1(u)= -\frac{1}{m+1}\ln \left(-\sqrt{\frac{m+1}{m-1}}(u-u_*)\right),~~u<u_*,
\end{equation}

\begin{equation}\label{C1=0b}
\nu_2(u)= -\frac{1}{m+1}\ln \left(\sqrt{\frac{m+1}{m-1}}(u-u_*)\right),~~u>u_*,
\end{equation}
where $u_*$ is an integration constant.
Let us note that for this solution we have
\begin{equation}\label{K-m}
K\equiv\frac{m+1}{m-1}>0,~~m\in (-\infty, -1) \& (1,\infty).
\end{equation}

From \eref{30-nu} we obtain
\begin{equation}\label{f2K-0}
f_2(\phi)=\frac{1}{\varkappa K_2 (m-1)}e^{-2m\nu}.
\end{equation}
Substituting the solution \eref{C1=0a} and  \eref{C1=0b} in \eref{f2K-0} one can find

\begin{equation}\label{f2K-01}
f_2(\phi)=\left(\varkappa K_2 (m-1)\right)^{-1}
\left[ \left(\frac{m+1}{m-1}\right)(u-u_*)^2\right]^{\frac{m}{m+1}}.
\end{equation}

Substituting the solution \eref{C1=0a} and  \eref{C1=0b} in \eref{Xphi} one can find

\begin{equation}
X(\phi)(\phi')^2=\frac{m}{\varkappa (m+1)^2 (u-u_*)^2}.
\end{equation}

Such relation gives for us possibility to find canonical or phantom scalar field when kinetic function $X(\phi)=const=\pm 1$.
In our case, when right hand side is always positive, we have only one choice $X=1$ and $\phi$ is a phantom field. The solution for $\phi$ is
\begin{equation}\label{phi-pha}
\phi=\pm\sqrt{\frac{m}{\varkappa}}\frac{1}{m+1}
\ln |u-u_*|+\phi_*.
\end{equation}
Inversed the solution \eref{phi-pha} and substituting the result in \eref{f2K-01} one can find
\begin{equation}\label{f2K-pp}
f_2(\phi)=\left(\varkappa K_2 (m-1)\right)^{-1}
 \left(\frac{m+1}{m-1}\right)^{\frac{m}{m+1}}
 e^{\pm 2 m (\phi-\phi_*)\sqrt{\frac{\varkappa}{m}}}.
\end{equation}

From the presentation \eref{f2-def} it is clear that the function $f_2(\phi)$ gives, in some sense, the deviation from GR. Thus we can see from \eref{f2K-pp} functional addition, keeping in mind the exchange $\phi \rightarrow R$.

\subsubsection{Asymptotes for $C_1=0$}

Let us note that from ansatz \eref{anz1} we can find
\begin{equation}\label{119-821}
e^{2\beta}=\left[e^{2\nu}\right]^m,~~
e^{2\lambda}=\left[e^{2\nu}\right]^{2m+1}.
\end{equation}

The solution for the case $C_1=0$ is described by

\begin{equation}\label{met_0}
e^{2\nu}=\left[ \left(\frac{m+1}{m-1}\right)(u-u_*)^2\right]^{-\frac{1}{m+1}},
~~e^{2\beta}=\left[e^{2\nu}\right]^m,~~
e^{2\lambda}=\left[e^{2\nu}\right]^{2m+1}.
\end{equation}

Let us consider candidates for horizons. First,  consider asymptotes when $u-u_* \rightarrow 0$.

Then, if
$ -\infty <m<-1$ we have

\begin{equation}\label{C_1=0-1}
e^{2\nu}\rightarrow 0,~~e^{2\beta}\rightarrow \infty,~~e^{2\lambda}\rightarrow \infty, ~~m\in (-\infty,-1).
\end{equation}

For all others values of $m$ ($m>1$, according to \eref{K-m}) we get $e^{2\nu}\rightarrow \infty $ and we have no candidates for horizons.

Let us consider this asymptote
in accordance with Bronnikov et al \cite{Bronnikov97} classification (see, \ref{CRIT} and \ref{AKRET}).

Let us represent main formulae for analysis.
According to the criterium ${C3}$ we must investigate the integral
$$
t^*=\int e^{2\beta}du=\int \left(K (u-u_*)^2\right)^{-\frac{m}{m+1}}du=
-\left(\frac{m+1}{m-1}\right)^{\frac{1}{m+1}}
(u-u_*)^{\frac{1-m}{m+1}}+const.
$$

Thus, one can see, that for the solution \eref{met_0}-\eref{C_1=0-1}, $t^* \rightarrow \infty $ always when $u \rightarrow u_*,~~ m\in (-\infty, -1)$.

The coefficient $\kappa$ for Hawking temperature of a surface $u=u^*$
for the solutions \eref{met_0} is
$$
\kappa = \lim_{u \rightarrow u^*} \left[\left( K\right)^{\frac{m}{m+1}} (u-u_*)^{\frac{m-1}{m+1}}\frac{1}{|m+1|}\right].
$$

The Kretschmann scalar can be calculated with \eref{Kret}-\eref{Kret-4}. For the solution \eref{met_0} we have

\begin{equation}\label{Kret-1-c0}
 \mathscr{K}_1=K^{\frac{2m+1}{m+1}}\frac{(1-m)}{(m+1)^2}
 (u-u_*)^{\frac{2m}{m+1}},
\end{equation}

\begin{equation}\label{Kret-2-c0}
\mathscr{K}_2=K^{\frac{2m+1}{m+1}}\frac{m}{(m+1)^2}
 (u-u_*)^{\frac{2m}{m+1}},
\end{equation}

\begin{equation}\label{Kret-3-c0}
\mathscr{K}_3=0,
\end{equation}

\begin{equation}\label{Kret-4-c0}
\mathscr{K}_4=
K^{\frac{m}{m+1}}\left(\frac{m^2-m+1}{1-m}\right) (u-u_*)^{\frac{2m}{m+1}}.
\end{equation}

The result for the case \eref{119-821}-\eref{met_0} is represented in Table \ref{tab1}. Remark that at $u \rightarrow \pm \infty$ there are a curvature singularities.
Thus we have a cold black hole.

\begin{table}
\begin{center}
\begin{tabular}{ |c|c|c|c|c|c|c|c|c|c| }
 \hline
$m$ &$e^{2\nu}$ &$e^{2\beta}$ &$e^{2\lambda}$ & $u-u_*$ & ${\bf C1}$ & ${\bf C2}$& ${\bf C3}$& ${\bf C4}$& ${\bf C5}$ \\
  \hline
 $ m<-1$ & $0$& $\infty$ & $\infty$ & $0$ &Yes & No &Yes & Yes &Yes \\
  \hline
 $m>1$ & $0$& $0$ & $0$ & $\infty$ & Yes & Yes &No & No &No \\
    \hline
   \end{tabular}
\end{center}
\caption{\label{tab1}The correspondence to horizon criteria for solutions \eref{119-821}-\eref{met_0}.}
\end{table}

\subsection{The solution $C_1=\mu^2>0$}\label{sbs6.1}
The solution
\begin{equation}\label{u-C1=mu}
u-u_*=\mp \frac{1}{(m+1)|\mu|}\ln \left[ \frac{2\mu^2+2|\mu|\sqrt{\frac{z^2}{m^{2}-1}+\mu^2}}{z}\right],~~
z=e^{(m+1)\nu},
\end{equation}
must be inverted in order to find the solution for $\nu (u)$. From \eref{u-C1=mu} one can obtain

\begin{equation}
\nu=\mu u -\frac{1}{m+1}\ln \left( K+\frac{1}{\mu^2}e^{2\mu(m+1)u}\right).
\end{equation}

It is clear that for positive $K \equiv \frac{m+1}{m-1}$ the logarithm's argument will  always be positive. The case when $K<0$, that is $-1<m<1$, and $K\in (-\infty, 0)$ implies a limited range for $u$. Therefore we will consider only the case when $K>0$.

Further, from \eref{Xphi} we find

\begin{equation}\label{Xphi-1}
2\varkappa X(\phi)(\phi')^2= \left[\frac{2m}{m^2-1}\right] \frac{\mu^2 E_1(u)}{\left(\frac{\mu^2}{m^2-1}+E_1(u)\right)^2}-
2\mu^2 m(m+2),
\end{equation}
where
$$
E_1(u)=e^{2\mu(m+1)u}.
$$

Thus we can state that for each given $\phi(u)\ne const.$ there are exist  $X(\phi)$, obtained from  \eref{Xphi-1}  and which gives the exact solution with
\begin{equation}
f_2(\phi (u))=\frac{1}{\varkappa K_2 (m-1)}e^{-2m\mu u}\left( \frac{1}{m^2-1}+\frac{1}{\mu^2}E_1(u)\right)^{\frac{2m}{m+1}}.
\end{equation}

Inverting the dependence $\phi $ on $u$ as $u(\phi)$ the dependence $f_2$ on $\phi$ will be restored. The kinetic function $X(\phi)$ can be defined by algorithm described in Sec. \ref{ss8-1}.

We may find the solution of \eref{Xphi-1} for canonical or phantom field if we set $X=\mp 1$. But in general case to evalute integral is rather difficult task. Therefore let us consider the special case when $m=-2$ and the last term in  \eref{Xphi-1} vanish.

The phantom scalar field is
\begin{equation}
\phi (u)=\mp \sqrt{\frac{7}{6 \varkappa}}\tanh^{-1}\left(\frac{1}{\sqrt{\frac{\mu^2}{3}e^{2\mu u}+1}}\right).
\end{equation}

\subsubsection{Asymptotes for $C_1=\mu^2>0$}

The solution for the case $C_1=\mu^2>0$ is described by
\begin{equation}
\label{C1=1MB}
e^{2\nu}=e^{2\mu u}\left[ K+\frac{1}{\mu^2}e^{2\mu (m+1)u} \right]^{-\frac{2}{m+1}},
~~e^{2\beta}=\left[e^{2\nu}\right]^m,
\end{equation}
\begin{equation}
e^{2\lambda}=\left[e^{2\nu}\right]^{2m+1},~~K=\frac{m+1}{m-1}.
\end{equation}

Let us rewrite the metric component $e^{2\nu}$
\begin{equation}\label{C1=1}
e^{2\nu}=\left[K e^{-\mu (m+1)u} +\frac{1}{\mu^2}e^{\mu (m+1)u} \right]^{-\frac{2}{m+1}},
\end{equation}

For this solutions horizons are not possible. If $u \rightarrow \pm \infty,$ then $ e^{-2\nu} \rightarrow \infty$ and Kretschmann scalar tends to $\infty$. If $u \rightarrow u_*=const.$ we obtain a regular solution for any $u_*$, including $u_*=0$. The results for these cases are presented in Table \ref{tab2} and Table \ref{tab3}.

\begin{table}
\begin{center}
\begin{tabular}{ |c|c|c|c|c|c|c|c|c| }
 \hline
$m$ &$e^{2\nu}$ &$e^{2\beta}$ &$e^{2\lambda}$  & ${\bf C1}$ & ${\bf C2}$& ${\bf C3}$& ${\bf C4}$& ${\bf C5}$ \\
  \hline
 $1<m<\infty$ & $0$& $0$ & $0$  & Yes & Yes &No & No &No \\
    \hline
   \end{tabular}
\end{center}
\caption{\label{tab2} The correspondence to horizon criteria for solutions \eref{C1=1MB}-\eref{C1=1} with $u \rightarrow +\infty$.}
\end{table}

\begin{table}
\begin{center}
\begin{tabular}{ |c|c|c|c|c|c|c|c|c| }
 \hline
$m$ &$e^{2\nu}$ &$e^{2\beta}$ &$e^{2\lambda}$  & ${\bf C1}$ & ${\bf C2}$& ${\bf C3}$& ${\bf C4}$& ${\bf C5}$ \\
  \hline
 $1<m<\infty$ & $0$& $0$ & $0$  & Yes & Yes &No & No &No \\
    \hline
   \end{tabular}
\end{center}
\caption{\label{tab3}The correspondence to horizon criteria for solutions \eref{C1=1MB}-\eref{C1=1} with  $u \rightarrow -\infty$.}
\end{table}

\subsection{The solution $C_1=-\alpha^2<0$}
The solution
\begin{equation}\label{u-C1-al}
u-u_*=\mp \frac{1}{\alpha(m+1)} \arctan\left(\frac{1}{\sqrt{\frac{z^{2}}{m^{2}-1}+\alpha^2}}\right),~~z=e^{(m+1)\nu},
\end{equation}
must be inverted in order to find the
solution for $\nu (u)$. From \eref{u-C1-al} one can obtain

\begin{equation}
\nu (u)=\frac{1}{m+1}\ln \left[K\alpha^2\left(1 + \tan^2(\alpha(m+1)u)\right)\right].
\end{equation}

We can obtain$f_2(\phi(u))$ from \eref{f-2-nu}

\begin{equation}
f_2(\phi (u))=\left[\alpha^2 (m-1)\left(1 + \tan^2(\alpha (m+1) u)\right)\right]^{-\frac{2m}{m+1}}\left(\varkappa K_2 (m-1)\right)^{-1}.
\end{equation}

From \eref{Xphi} one can find
\begin{equation}
\nonumber
\varkappa X(\phi)(\phi')^2=\alpha^4 m(m^2-1)\left[ 1 + \tan^2(\alpha (m+1) u)\right]^2+
\alpha^2 m(m+2).
\end{equation}

Inverting the dependence $\phi $ on $u$ as $u(\phi)$ the dependence $f_2$ on $\phi$ will be restored. The kinetic function $X(\phi)$ can be defined by algorithm described in Sec. \ref{sbs6.1}

Once again, we consider the case when $m=-2$. The solution for phantom field with $X=1$ is
\begin{equation}
\phi (u)=\pm \alpha^2 \sqrt{\frac{21}{2\varkappa}}\left[(u-u_*) + \frac{1}{\alpha}\tan (\alpha u) \right] +\phi_*.
\end{equation}

Thus, we can obtain the following relation for $f_2(\phi)$:
\begin{equation}
f_2(\phi)= \frac{\alpha^2}{\varkappa K_2}\left( 1+\frac{2\varkappa}{21 \alpha^2}(\phi-\phi_*)\right)^{-4}.
\end{equation}

\subsubsection{Asymptotes for  $C_1=-\alpha^2<0$.}\label{regular}

The solution for the case $C_1=-\alpha^2<0$ is described by
\begin{equation}\label{e-nu-al}
e^{2\nu}=\left[K\alpha^2\left(1 + \tan^2 (\alpha (m+1)u)\right) \right]^{\frac{2}{m+1}},
~~e^{2\beta}=\left[e^{2\nu}\right]^m,~
e^{2\lambda}=\left[e^{2\nu}\right]^{2m+1}.
\end{equation}

Horizons are possible when $u \rightarrow \pm \frac{\pi}{2\alpha (m+1)}$. If $u \ne \pm \frac{\pi}{2\alpha (m+1)}$ we have finite value of $e^{-2\nu}$ and Kretschmann scalar is finite everywhere.

 If $u \rightarrow \pm \frac{\pi}{2\alpha (m+1)}$ we have the following asymptotes:
\begin{itemize}
\item
if
$ -\infty <m<-1$ we have
$$ e^{2\nu}\rightarrow 0,~~e^{2\beta}\rightarrow \infty,~~e^{2\lambda}\rightarrow \infty, ~~K=\frac{m+1}{m-1}>0.
$$

\end{itemize}

\begin{table}
\begin{center}
\begin{tabular}{ |c|c|c|c|c|c|c|c|c| }
 \hline
$m$ &$e^{2\nu}$ &$e^{2\beta}$ &$e^{2\lambda}$  & ${\bf C1}$ & ${\bf C2}$& ${\bf C3}$& ${\bf C4}$& ${\bf C5}$ \\
  \hline
 $-\infty<m<-1$ & $0$& $\infty$ & $\infty$  & Yes & No &Yes & Yes & Yes \\
  \hline
\end{tabular}
\end{center}
\caption{\label{tab4}The correspondence to horizon criteria for solutions \eref{e-nu-al} with $u \rightarrow \pm \frac{\pi}{2\alpha (m+1)}$.}
\end{table}

The Hawking temperature can be obtained with the following limits for parameter $\kappa$:

\begin{equation}\label{up-kap}
\kappa =\lim_{u \rightarrow u^*} \left[|2\alpha| \left(K\alpha^2\right)^{-\frac{2m}{m+1}}
\left[ \cos (\alpha (m+1)u)\right]^{\frac{4m}{m+1}}|\tan (\alpha (m+1)u)|\right].
\end{equation}

The criteria are represented in Table \ref{tab4}. Thus, in principle, we have cold black hole. But this cold black hole belongs to the second type discussed in \cite{kb1} since the horizons are at an infinite geodesic distance. However, differently from what was found in \cite{kb1}, both asymptotic correspond to horizons: the solution is not asymptotical flat. In brief, it is a regular solution which has some similarities with a wormhole but with horizons at the spatial infinities. The spacetime is geodesically complete.

\subsubsection{Searching for wormhole solution}

It is clear that one can choose two value of generalised radial coordinate $u$ as two necessary surfaces: $u=u_1=-\frac{\pi}{2\alpha (m+1)}$ and $u=u_2=\frac{\pi}{2\alpha (m+1)}$. Then the criteria {B1, B2, B3} are true, but {B4} is not true. Asymptotically we have $e^{-\beta-\nu}\beta' \rightarrow 0$. The circumference-radius ratio tends to zero. This is impossible, so we have no wormhole solution.

\section{Exceptional solution $\beta=-\nu$ }\label{EXS}

In previous section we excluded the case $m=-1$. Let us study this special case. Equations \eref{30-nu-az}-\eref{34-phi-az} are reduced to the following ones

\begin{equation}\label{30-nu-ex}
e^{2\nu}\nu''=\varkappa K_2 f_2(\phi),
\end{equation}

\begin{equation}\label{31-X-ex}
2\nu''-2(\nu')^2=\varkappa \left(-2X(\phi)(\phi')^2-e^{-2\nu} K_2 f_2(\phi)\right),
\end{equation}

\begin{equation}\label{32-beta-ex}
1+\nu''=-\varkappa e^{-2\nu}K_2 f_2(\phi),
\end{equation}

\begin{equation}\label{34-phi-ex}
2X(\phi)\phi''+X'_\phi (\phi')^2-e^{-2\nu}K_2f_{2,\phi}=0.
\end{equation}

Substituting $\varkappa K_2 f_2(\phi)$ from \eref{30-nu-ex} into rhs of \eref{32-beta-ex} one can find the solution
\begin{equation}\label{-1/2}
\nu''=-\frac{1}{2},~~\nu'=-\frac{u}{2}+c_1,~~\nu= -\frac{u^2}{4}+c_1 u +c_2.
\end{equation}

Substituting solution \eref{-1/2}  into \eref{31-X-ex} we find
\begin{equation}\label{1/2-phi-ex}
X(\phi)(\phi')^2 = \frac{1}{2\varkappa}\left[ \frac{3}{2}+2\tilde{u}^2\right],~~\tilde{u}= c_1 - \frac{u}{2}.
\end{equation}

Choosing $X=1$ we can find the solution for phantom scalar field
\begin{equation}\label{102}
\phi(u)=\mp \frac{1}{4\sqrt{\varkappa}}\left[ \frac{2\tilde{u}}{\sqrt{3}} \sqrt{1+\frac{4}{3}\tilde{u}^2} +\sinh^{-1}\left( \frac{2\tilde{u}}{\sqrt{3}} \right)\right]+\phi_*,
\end{equation}
where $\sinh^{-1}(x)$ means ${\rm arcsinh}(x)$.

From equation \eref{34-phi-ex} for the case $X=1$ one can obtain the relation
\begin{equation}\label{c3}
(\phi')^2=-\frac{\nu}{\varkappa}+c_3.
\end{equation}

Using \eref{1/2-phi-ex}, \eref{-1/2} and \eref{c3} one can obtain the restriction on the constants
\begin{equation}\label{cnsts}
\varkappa c_3-c_2=\frac{3}{4}c_1^2.
\end{equation}

Thus we have the solution for phantom field in the exceptional case $m=-1$ \eref{-1/2}, \eref{102}  with \eref{cnsts}.

Also, one can obtain the remaining functions for this case as $\beta=\lambda=-\nu$, and corresponding metric is
\begin{equation}\label{SphSy_mbr_min1}
  ds^2=-e^{2\nu (u)}dt^2+e^{-2\nu (u)} du^2 +e^{-2\nu (u)} \left(d\theta^2+\sin^2 \theta d\varphi^2 \right),
\end{equation}
where $\nu=-\frac{u^2}{4}+c_1 u +c_2$.
Defining $\sqrt{2}x = u - 2c_1$ and $\bar k = c_1^2 + c_2$, the metric can be written as,
\begin{eqnarray}
ds^2 = - e^{2\bar k}e^{-x^2}dt^2 + 2e^{-2\bar k}e^{x^2}dx^2 + e^{-2\bar k}e^{x^2} \left(d\theta^2+\sin^2 \theta d\varphi^2 \right).
\end{eqnarray}
It is clear that there are possible horizons at $x \rightarrow \pm \infty $ and
$$
e^{2\nu}\rightarrow 0,~~e^{2\beta}\rightarrow \infty,~~e^{2\lambda}\rightarrow \infty.
$$
This solution gives cold black hole similar to the case discussed in subsection \ref{regular}.

\subsubsection*{Searching for wormhole solution}

It is clear that one can choose two value of generalised radial coordinate $u$ as two necessary surfaces: $u=u_{-\infty}=-\infty $ and $u=u_{\infty}=\infty $. Then the criteria {B1, B2, B3} are true, but {B4} is not true. Asymptotically we have $e^{-\beta-\nu}\beta' \rightarrow \pm \infty $.  The circumference-radius ratio tends to infinity. As one of the possibilities we can imagine $r \to 0$, that is a string space.  So we have no wormhole solution.

\section*{Conclusions}

The main idea to simplify investigation of black holes and wormholes in kinetic scalar curvature extended $f(R)$ gravity is connected with the presentation of mentioned model in the form of CSGM in the framework of Einstein gravity. The CSGM in the case with $f(R)=f_1(R)+X(R)\nabla_\mu R \nabla^\mu R$ has the fixed form of the potential and the chiral metric components. Because we suggested to exploit rather new theory we remind detail of CSGM, namely the form of the action, energy-momentum tensor, gravitational and chiral fields equation. Also we presented the main features of transformation of $f(R,(\nabla R)^2)$ gravity
to Einstein gravity with additional scalar fields which described by CSGM. Further we investigate the model in spherically symmetric static spacetime in harmonic coordinates.

To simplify models equations we consider special scaling transformation when non-minimally coupled to gravity model converted to Einstein frame by virtue of Weyl unit conformal function.
We study in this article only the special case when $ \chi= -\sqrt{\frac{3}{2}}\ln 2$. It is difficult to find exact solutions for the case when $ \chi \ne -\sqrt{\frac{3}{2}}\ln 2$. Nevertheless we can use obtained solutions for searching solutions when $\chi $ is a function on a coordinate $u$.
After that, we study quasi-GR solutions of the model when $f_1(\phi)=\phi/2$ and consequently the potential $ W(\phi)=0$. In this case we obtained two classes of three and five parametric solutions. Investigation of a geometry of obtained solutions gives various types of them. Namely, cold black holes, wormholes solutions with two minkovskian asymptotic, solution with Schwarzschild-type horizon, solutions with singular center and a naked singularity. The quasi-GR solutions we compare to the similar solutions obtained earlier in Brans-Dicke gravity.

New solutions for the metric components and explanation of kinetic function $X(\phi)$ derivation we were able to find under introduction of the special relation between metric components (ansatz) $\beta = m \nu,~ m=const.$
Such relation allows us to reduce the system of gravitational and field equations to one non-linear equation of the first order on metric exponent $\nu (u)$ and to derive the equation for definition of kinetic function and squared derivative of the scalar field combination  $X(\phi)(\phi')^2$.

It was found three classes of solutions in accordance with an integration constant $C_1 $ value. A geometry of all classes of solutions were investigated. Once again we found various types of solution, including cold black holes. For each class of solutions we define canonical (phantom) field by virtue of setting $X(\phi)=1.$

Thus, within the framework of the proposed approach, we have the ability to describe astrophysical objects based on modified  gravity theories under consideration, which were considered earlier in a cosmological context \cite{cfo-Chervon:2018ths,cfo-Tsoukalas2017,Chervon:2019jfu}.

The solutions described in this article depend on the configuration of the scalar field. One may ask is if such fact implies a violation of the no-hair conjecture.
In principle the presence of a scalar field which determine a specific black hole type structure may characterises a hair. But, a final answer to the possible violation of the no-hair conjecture may be more complex: even if in principle it is possible to try to identify a charge related to the the scalar field, this charge must be associated with a conserved current.  Moreover, in the present case there is another subtle aspect of the problem: the original model is purely geometric and the scalar field is defined from the geometric quantities. Hence, these
considerations point out to the necessity of a specific and deep investigation if the scalar field in the analysis carried out here may allow to state that there is a violation of the no-hair conjecture.

An important task for possible further studies of the application of $f(R)$ gravity with a kinetic scalar of curvature to the description of astrophysical objects is the calculation of the spectra of gravitational waves. After experimental estimates of the propagation velocity of gravitational waves, which turned out to be very close to the speed of light in a vacuum (up to $10^{-16}$) \cite{GBM:2017lvd}, a description of astrophysical objects based on modified gravity theories should meet this result \cite{Ezquiaga:2017ekz}.
Thus, the calculation of the characteristics of gravitational waves in the constructed models of astrophysical objects is an important area of our further research.

\section*{Acknowledgments}

S.V. Chervon is thankful CNPq (Brazil) for support  of his visit to Brazil, Vitoria, UFES, where the work was finalised.
  S.V. Chervon and I.V. Fomin note that the study has been carried out under the financial support of RFBR by grant 20-02-00280 a. S.V. Chervon is grateful for the support of the Program of Competitive Growth of Kazan Federal University. J.C. Fabris thanks CNPq (Brazil) and FAPES (Brazil) for partial financial support.

\begin{appendix}

\section{Hawking temperature and Kretschmann scalar in harmonic coordinates}\label{HKRET}

In the work \cite{Bronnikov97} it was proved that an infinite Hawking
temperature indicates that an assumed horizon exhibits a singularity.

The Hawking temperature of a surface $u=u^*$ is defined as \cite{Wald-84}
\begin{equation}\label{T-H-gen}
T_H=\frac{\kappa}{2\pi},~~\kappa = \lim_{u \rightarrow u^*} \left(e^{-2\beta}|\nu'|\right),
\end{equation}
where the Boltzmann constant $k_B$ and the Planck constant $\hbar$ are set equal to 1. Also it was suggested that both functions $\nu (u)$ and $e^{\nu - \lambda}\nu'$ are monotonic in some neighbourhood of $u^*$.

The Kretschmann scalar in harmonic coordinates can be represented as
\begin{equation}\label{Kret-hc}
 \mathscr{K} =4 \mathscr{K} _1^2+8 \mathscr{K} _2^2+8\mathscr{K} _3^2+4 \mathscr{K}_4^2,
\end{equation}
where
\begin{equation}\label{Kret-1hc}
 \mathscr{K}_1=R^{01}_{~~01}=-e^{-2(2\beta+\nu)}
 \left(\nu'' -2\beta'\nu'\right),
\end{equation}

\begin{equation}\label{Kret-2hc}
\mathscr{K}_2=R^{02}_{~~02}=R^{03}_{~~03}= e^{-2(2\beta+\nu)}\beta' \nu',
\end{equation}

\begin{equation}\label{Kret-3hc}
\mathscr{K}_3=R^{12}_{~~12}=R^{13}_{~~13}= e^{-2(2\beta+\nu)}\left( \nu''-\beta'\nu'-(\nu')^2\right),
\end{equation}

\begin{equation}\label{Kret-4hc}
\mathscr{K}_4=R^{23}_{~~23}=-e^{-2\beta}+ e^{-2(2\beta+\nu)}(\beta')^2.
\end{equation}

Thus, to calculate the Kretschmann scalar it needs to insert the functions and derivatives of the solutions represented in subsection \ref{ex_sols} into (\ref{Kret-hc})-(\ref{Kret-4hc}).

\section{Criteria for black hole and wormhole selection}\label{CRIT}

 In the work by Bronnikov \cite{kb-96-GC} it was suggested the criteria for searching horizons for BH and wormhole in the case of multidimensional generalized scalar-tensor gravity. The criteria when metric describes BH, i.e. when the metric possesses an event horizons was formulated and named as A1-A4.
 Later, in the work \cite{Bronnikov97} similar criteria were applied for 4D spherically symmetric static spacetime in Einstein gravity.
We will follow by criteria for black holes selection
 in accordance with Bronnikov et al classification \cite{Bronnikov97}.
Namely, black hole  solutions with the metric
 (\ref{SphSy_mbr}) are conventionally selected by the following criteria: at some surface $u=u^*$ (horizon)
\begin{itemize}
\item {C1.} $ e^\nu \rightarrow 0 $ (the timelike Killing vector becomes null);
\item {C2.} $ e^\beta$ is finite (finite horizon area)\footnote{In fact, a cold black hole has an infinite area
    \cite{Campanelli1993}.};
\item {C3.} The integral $t^*=\int e^{\lambda - \nu}du \rightarrow \infty$ as $ u \rightarrow u^*$ (invisibility of the horizon for an observer at rest);
 \item {C4.}  The Hawking temperature $T_H$ is finite;
 \item {C5.} The Kretschmann scalar $\mathscr{K}$ is finite at $ u = u^*$.

\end{itemize}

We have applied this classification to the solutions studied in this article.

The criterium for searching of wormholes in \cite{kb-96-GC} can be reduced for 4D spherically symmetric static spacetime in Einstein gravity by the following manner. The solution of the metric (\ref{SphSy_mbr}) describes a (static, traversable) wormhole if at $u=u^*$ and at some other value of the radial coordinate $u=u_{\infty}$
\begin{itemize}
\item {B1.} $e^\nu$ remains finite;
\item {B2.} $e^\beta \rightarrow \infty$;
\item {B3.} There is an infinite path along the radius, i.e., the integral $ \int e^\lambda du$ diverges;
\item {B4.} A correct flat-space circumference-radius ratio for
coordinate circles is asymptotically valid, i.e., $e^{-\beta-\nu} \beta' \rightarrow 1$.
\end{itemize}


Thus, one can use these criteria to analyze the static solutions of gravitational field’s equations for the correspondence of black holes and wormholes solutions.

\section{Hawking temperature and Kretschmann scalar in harmonic coordinates with ansatz}\label{AKRET}

The Hawking temperature of a surface $u=u^*$ is defined as
$$
T_H=\frac{\kappa}{2\pi},~~\kappa = \lim_{u \rightarrow u^*} \left(e^{-2m\nu}|(m^2-1)^{-1}e^{2(m+1)\nu}+
C_1|^{1/2}\right),
$$
where the Boltzmann constant $k_B$ and the Planck constant $\hbar$ equal to 1. Also it was suggested that both functions $\nu (u)$ and $e^{\nu - \lambda}\nu'$
are monotonic in some neighbourhood of $u^*$.

Taking into account ansatz solutions \eref{dd-nu}, \eref{nu-m1} the Kretschmann scalar can be calculated with

\begin{equation}\label{Kret}
 \mathscr{K} =4 \mathscr{K} _1^2+8 \mathscr{K} _2^2+8\mathscr{K} _3^2+4 \mathscr{K}_4^2,
\end{equation}
where
\begin{equation}\label{Kret-1}
 \mathscr{K}_1=-C_1 e^{-2\nu (2m+1)}+(m+1)^{-1}e^{-2m\nu},
\end{equation}

\begin{equation}\label{Kret-2}
\mathscr{K}_2=m (m^2-1)^{-1}e^{-2m\nu} +C_1 e^{-2(2m+1)\nu},
\end{equation}

\begin{equation}\label{Kret-3}
\mathscr{K}_3=m (m+1)C_1 e^{-2(2m+1)\nu},
\end{equation}

\begin{equation}\label{Kret-4}
\mathscr{K}_4=(1-m^2)^{-1}e^{-2m\nu} -m^2 C_1 e^{-2(2m+1)\nu}.
\end{equation}

Thus, one can use these expressions for the Kretschman scalar to analyze the correspondence of the solutions obtained to the black holes and wormholes criteria.

\end{appendix}

\end{document}